\RequirePackage[2020-02-02]{latexrelease}
\documentclass[aps, 12pt, final, notitlepage, oneside, onecolumn, nobibnotes, nofootinbib, superscriptaddress, noshowpacs, centertags]{revtex4}

\usepackage{amsmath,amsthm,amssymb}
\usepackage{mathtext}
\usepackage[T1, T2A]{fontenc}
\usepackage[english]{babel}
\usepackage[usenames]{color}
\usepackage{multirow}

\usepackage{graphicx}

\begin{document}


\title{MHD Modeling of the Molecular Filament Evolution}

\author{\firstname{I.~M.}~\surname{Sultanov}}
\email{syltahof@yandex.ru}
\affiliation{%
Chelyabinsk State University, Chelyabinsk, Russia
}
\author{\firstname{S.~A.}~\surname{Khaibrakhmanov}}
\affiliation{%
Saint Petersburg State University, St. Petersburg, Russia
}%
\affiliation{%
Chelyabinsk State University, Chelyabinsk, Russia
}%
\affiliation{%
Ural Federal University, Yekaterinburg, Russia
}

\begin{abstract}
We perform numerical magnetohydrodynamic (MHD) simulations of the gravitational collapse and fragmentation of a cylindrical molecular cloud with the help of the FLASH code. The cloud collapses rapidly along its radius without any signs of fragmentation in the simulations without magnetic field. The radial collapse of the cloud is stopped by the magnetic pressure gradient in the simulations with parallel magnetic field. Cores with high density form at the cloud's ends during further evolution. The core densities are $n \approx 1.7 \cdot 10^8$~and $2 \cdot 10^7$~cm$^{-3}$ in the cases with initial magnetic field strengths $B = 1.9 \cdot 10^{-4}$~and $6 \cdot 10^{-4}$ G, respectively. The cores move toward the cloud's center with supersonic speeds $|v_{z}|=3.6$~and $5.3$~km$\cdot$s$^{-1}$. The sizes of the cores along the filaments radius and filament's main axis are 
$d_{r} = 0.0075$~pc and $d_{z} = 0.025$~pc, $d_{r} = 0.03$~pc and $d_{z} = 0.025$~pc, respectively. The masses of the cores increase during the filament evolution and lie in range of $\approx 10-20\,M_\odot$. According to our results, the cores observed at the edges of molecular filaments can be a result of the filament evolution with parallel magnetic field.

Keywords: magnetic fields, magnetohydrodynamics (MHD), methods: numerical, ISM: clouds

\end{abstract}

\maketitle

\section*{Introduction}
\label{sect:intro}

Modern observations show that interstellar clouds have a filamentary structure, which is traced from HI super clouds to individual molecular clouds~\cite{andre2014}. Filaments appear as elongated structures in the radiation maps of the interstellar medium. Such filaments can be either cylindrical clouds or edge-on sheets of molecular gas~\cite{DK2017}. Most protostellar clouds in which star formation occurs are located within filamentary molecular clouds~\cite{konyves2015}. Therefore, the study of the structure and evolution of molecular filaments is important for the theory of star formation.

Observations show that the typical width of molecular filaments is of $10^{-1}$~pc, and their length varies from several pc to hundreds of pc. The temperature in the filaments ranges from $10$~to $25$~K, and the gas density ranges from $10^4$~up to $10^5$~cm$^{-3}$~\cite{DK2017}.

Polarization mapping of molecular clouds revealed the presence of large-scale magnetic fields~\cite{ward_thompson2017}. The magnetic field is usually aligned with the main axis of the filament in low density clouds, and it is perpendicular to the cloud's axis in high density clouds. Measurements of the Zeeman splitting of OH lines and estimations with the Chandrasekhar-Fermi method showed that the strength of the magnetic field in the filaments increases with the column density and lies in the range from $10^{-5}$~G for low density clouds with $N=10^{19}$~cm$^{-2}$ to $10^{-3}$~G for the dense filaments with $N=10^{23}$~cm$^{-2}$.

The question of the nature of the filamentary structure of the interstellar medium is currently open~\cite{hacar2023}. Several mechanisms of the filament formation have been proposed in application to the various levels of the hierarchy of the interstellar medium: Parker and thermal instabilities on the scales of the spiral arms of the Galaxy, gravitational instability, collisions of interstellar shock waves in a turbulent medium, and large-scale anisotropic motions of the gas in the interstellar medium with magnetic field. The evolution of filaments after their formation depends on their initial
state and external conditions.

Gravitational focusing results in the formation of dense cores at the ends of isolated homogeneous isothermal filaments. This mechanism is called as "end-dominated collapse~\cite{Bastien}" or "edge fragmentation~\cite{hacar2023}". An example of such a cloud is the S242 filament, at the ends of which the cores with density of the order of $10^5$ cm$^{-3}$ and the size of $1$ pc are observed~\cite{dewanghan}.

Small longitudinal perturbations of the filaments can lead to the development of gravitational instability~\cite{ChFe, stod, ostriker}. In the case of filaments without magnetic field, instability develops for the perturbations with a wavelength 4 times larger than the radius of the homogeneous part of the filament. Instability leads to the formation of gravitational "sausages" and, subsequently, cores, which are distributed along the filament with a characteristic distance between them of the order of the fastest growing mode. The shape of the cores formed as a result of the gravitational fragmentation is close to spherical~\cite{IM}. The NGC 2024S/Orion B filament is an example of an object, in which signs of gravitational fragmentation are observed. Cores in this filament have sizes of the order of $10^{-2}$~pc and masses of $\sim 1 \ M_{\odot}$~\cite{shimajiri}. The velocity profile along the filament is periodic with wavelength of $\lambda\sim 0.2$~pc. The cores are spatially shifted relative to the velocity spikes by $\lambda/4$, which indicates the gravitational fragmentation of the filament. Another example of such a
filament is WB~673~\cite{ryabukhina2022}.

Simulations of the filament fragmentation with magnetic field are necessary for the interpretation of the observational data. Seifried and Walch~\cite{sw2015} used the numerical code FLASH to simulate the evolution
of filaments with turbulence and different magnetic field orientations. The authors identified several modes of filament fragmentation depending on the initial conditions: fragmentation at the ends, gravitational fragmentation of the filament with the subsequent formation of evenly distributed cores, and global collapse of the filament towards the center of the cloud. The authors showed that the pressure gradient of the parallel magnetic field maintains almost constant filament width of the order of 0.1 pc. Subsequently,
this conclusion was confirmed in the MHD simulations by Dudorov and Khaibrakhmanov~\cite{DK2017}.

In this work, numerical simulations of the homogeneous molecular filaments with parallel magnetic field is performed and the properties of the cores formed as a result of fragmentation at the edges of the
filament are determined. The main attention is paid to the influence of the magnetic field on the fragmentation, as well as on the internal structure, size and mass of the resulting cores.

Section~\ref{sect:model} describes the problem statement, the basic equations, and the numerical code FLASH, which is used to solve the equations. Subsection~\ref{sect:general_results} presents the results of the simulation of the filament evolution without magnetic field and with weak magnetic field. The results of the simulations with stronger magnetic field are given in Subsection~\ref{sect:mag_results}. Subsection~\ref{sect:cores} describes the characteristics of cores
formed in the simulations with magnetic field. In summary, we outline and discuss our main results.

\section{Model}
\label{sect:model}
\subsection{Problem Statement}
\label{sect:problem}
In this work, we simulate the gravitational collapse of a cylindrical molecular cloud (filament) with length $H_0 = 10$~pc and radius $r_0 = 0.2$~pc. The molecular weight of the gas is $\mu= 2.31$, temperature $T_0=10$~K, concentration $n_0 = 10^5$~cm$^{-3}$. The filament mass-per-length $M/L = 658 \ M_{\odot}\cdot$pc$^{-1}$ exceeds the critical value $(M/L)_{\rm crit} = 16.6 \ M_{\odot}\cdot$pc$^{-1}$~\cite{stod, ostriker}. Therefore, the filament is gravitationally unstable. Collapse simulations, which take into account radiation transfer, showed that the thermal energy of the compressed gas is effectively released and the gas temperature remains approximately constant in the concentration range $n = [10^{5}, \ 10^{11}]$~cm$^{-3}$. Therefore, for simplicity, we assume that the gas is characterized by the equation of state with the effective adiabatic index $\gamma = 1.001$ corresponding to the isothermal compression. The corresponding ratio of the cloud's thermal energy to the absolute value of its gravitational energy is  $\varepsilon_{\rm T} = 0.003$. The speed of sound in the filament is $c_{\rm s} = 0.19$~km$\cdot$s$^{-1}$.

To study the role of the magnetic field in the evolution of filaments, we performed three simulations with different ratios of the cloud's magnetic energy to the absolute value of its gravitational energy: $\varepsilon_{\rm m} = 0$~(run 'HD'), $\varepsilon_{\rm m} = 0.03$~(run 'MHD-1'), $\varepsilon_{\rm m} = 0.27$~(run 'MHD-2'). Corresponding magnetic field strengths are $B=0, \ 1.9 \cdot 10^{-4}, \ 6 \cdot 10^{-4}$~G, respectively. The magnetic field is parallel to the filament's axis in both cases. The filament is in pressure equilibrium with the external environment with concentration $n = 10^4$~cm$^{-3}$~and temperature $T = 100$~K. The free fall time for the
chosen density is $t_{\rm ff} \approx 10^5$~years.

Let us find out whether it is necessary to take into account the magnetic diffusion when studying the initial stages of the filament collapse. To do this, we need to estimate the magnetic Reynolds number,
\begin{equation}
   R_{\rm m} = \dfrac{v{_0} l{_0} }{\nu}, 
\end{equation}
where $v_0$ and $l_0$~are typical gas velocity and spatial scales, and $\nu$~is the magnetic diffusivity. We choose the filament radius as the $l_0$ and the characteristic speed as $v_0=\dfrac{r_0}{t_{\rm ff}}$.

Magnetic flux dissipation can be caused either by Ohmic dissipation (OD) or by magnetic ambipolar diffusion (MAD). We use the equations from~\cite{DK2014} to estimate the corresponding diffusivities

\begin{eqnarray}
\nu = \begin{cases} 480 x^{-1} T^{1/2}\,\,\mbox{cm}^2\cdot\mbox{s}^{-1} & \ (\text{OD}),  \\ \dfrac{B^2}{4 \pi x \rho^2 \eta_{\rm in}} & \ (\text{MAD}), \end{cases}
\end{eqnarray}

where $x$~is the ionization fraction, $T$~is the gas temperature, $B$~is the magnetic field strength, $\rho$~is the gas density, $\eta_{\rm in}=m_{\rm i}\langle\sigma v\rangle_{\rm in}/m_{\rm i}(m_{\rm i} + m_{\rm n})$~is the coefficient of the interaction between ions with mass $m_{\rm i}=30\,m_{H}$ and neutrals with mass $m_{\rm n}=\mu\,m_{H}$, where $\langle\sigma v\rangle_{\rm in}=2\cdot 10^{-9}\,\text{cm}^3\,\text{s}^{-1}$, $m_H$~is the mass of the hydrogen atom

In the considered range of densities, the ionization fraction can be estimated from the balance of the ionization by cosmic rays with rate $\xi$ and radiative recombinations: $x = \sqrt{\dfrac{\xi}{\alpha_{\rm r} n}}$, where $\alpha_{\rm r}=6.21\cdot10^{-11}\,T^{-1/2}\,\text{cm}^3\,\text{s}^{-1}$~is the coefficient of radiative recombinations, $n$~is the gas concentration~\cite{DS1987}.

Using typical parameters for the interstellar medium, we find for the case of Ohmic dissipation:
\begin{eqnarray}
    R_{\rm m}^{\mbox{OD}} = 1.7\cdot10^{14} \left(\dfrac{r_0}{0.2 \ pc}\right)^{2} \left(\dfrac{\xi}{10^{-17} \ s^{-1}}\right)^{1/2} \left(\dfrac{T_0}{10 \ K}\right)^{-1/4},\label{eq:RmOD}
\end{eqnarray}
for ambipolar diffusion:
\begin{eqnarray}
    R_{\rm m}^{\mbox{MAD}} = 6.6\cdot10^3 \left(\dfrac{r_0}{0.2 \ pc}\right)^{2} \left(\dfrac{B_0}{1.9\cdot10^{-4} \ G}\right)^{-2} \left(\dfrac{\xi}{10^{-17} \ s^{-1}}\right)^{1/2}\left(\dfrac{n_0}{10^5 \ cm^{-3}}\right)^2 \left(\dfrac{T_0}{10 \ K}\right)^{1/4}.\label{eq:RmAD}
\end{eqnarray}

Estimates~(\ref{eq:RmOD}) and (\ref{eq:RmAD}) show that $R_{\rm m} \gg 1$ for the parameters considered, that is, the magnetic field is frozen into the gas and the ideal MHD approximation is valid.

\subsection{Basic Equations and Solution Methods}
\label{sect:equations}
We study the evolution of the molecular filament using the system of ideal MHD equations:

\begin{eqnarray}
    \dfrac{\partial \rho}{\partial t} + \nabla (\rho \boldsymbol{v}) &=& 0, \label{eq:rho}\\
    \dfrac{\partial \boldsymbol{v}}{\partial t}+(\boldsymbol{v}\nabla)\boldsymbol{v} &=& -\dfrac{1}{\rho}\nabla P -\nabla \Phi - \dfrac{1}{4\pi \rho}\boldsymbol{B} \times (\nabla \times \boldsymbol{B}), \\
    \dfrac{\partial \boldsymbol {B}}{\partial t}  &=& \nabla \times (\boldsymbol{v} \times \boldsymbol{B}), \\
    \dfrac{\partial}{\partial t} \left[\rho\left(\varepsilon + \dfrac{v^2}{2} + \Phi\right) + \dfrac{B^2}{8\pi}\right] &=& -\nabla\left[\rho \boldsymbol{v}\left(\varepsilon + \dfrac{v^2}{2} + \dfrac{P}{\rho} + \Phi\right) + \dfrac{1}{4\pi}\boldsymbol{B} \times (\boldsymbol{v} \times \boldsymbol{B})\right], \label{eq:e}\\
    \Delta \Phi &=& 4 \pi G \rho, \\
    P &=& (\gamma - 1)\rho \varepsilon,
\end{eqnarray}
where $\rho$, $\boldsymbol{v}$ and $P$~are the density, velocity vector and pressure of the gas, ${\Phi}$~is the gravitational potential, $\boldsymbol{B}$~is the magnetic induction, $\varepsilon$~is the internal energy of the gas, $G$~is the gravitational constant.

To simulate the evolution of the filament, we use the numerical code FLASH 4, in which the adaptive mesh refinement (AMR) technology is implemented~\cite{Fryxell}. The equations of ideal MHD (\ref{eq:rho}--\ref{eq:e}) are solved using the Godunov-type MUSCL scheme~\cite{muscl}. We consider a 3D problem in Cartesian coordinates. The $z$-axis corresponds to the symmetry axis of the filament. The sizes of the computational domain in the $x$-, $y$-, and $z$-directions are $1.93 \times 1.93 \times 12.9$~pc$^{3}$, and 7 levels of the AMR grid are used. The sizes of the largest cell in the $x$-, $y$-, and $z$-directions are $0.24 \times 0.24 \times 1.61$~pc$^{3}$, the sizes of the smallest cells are $0.0037 \times 0.0037 \times 0.025$~pc$^{3}$, which corresponds to the effective grid resolution $512 \times 512 \times 512$ at the 7th AMR level. The Poisson's equation for gravity is solved using the multipole method based on the Barnes-Hut tree~\cite{BHtree}.

\section{Results}
\label{sect:results}

\subsection{General Picture of the Filament's Evolution}
\label{sect:general_results}

\begin{figure}[t]
\includegraphics[width=0.9\textwidth]{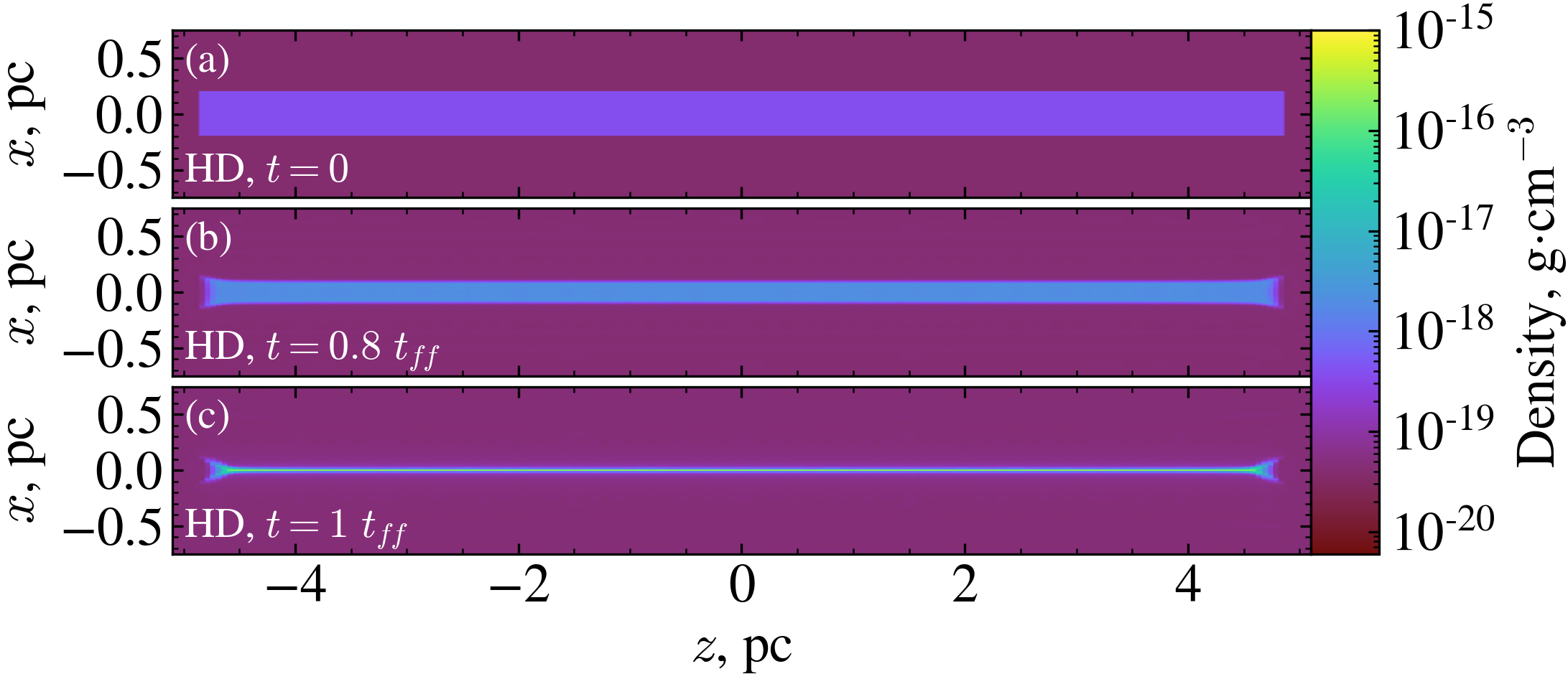}
\caption{The distribution of gas density in the $x-z$-plane for run HD at $t = 0$~(a), $t = 0.8 \, t_{\rm ff}$~(b), and $t = 1 \, t_{\rm ff}$~(c).}
\label{Fig. 1:}
\end{figure}

In Figure \ref{Fig. 1:}, we plot the gas density distribution in the $x-z$~plane for run 'HD' at $t = 0, \, 0.8 \, t_{\rm ff}, \, 1 \, t_{\rm ff}$. The simulations show that non-magnetic filament freely collapses along the radius and the density at the center of the filament increases by 3 orders of magnitude by the time $t=1 \, t_{\rm ff}$, the filament width along its radius is $0.004$~pc. The filament does not fragment.

\begin{figure}[h!]
\includegraphics[width=0.9\textwidth]{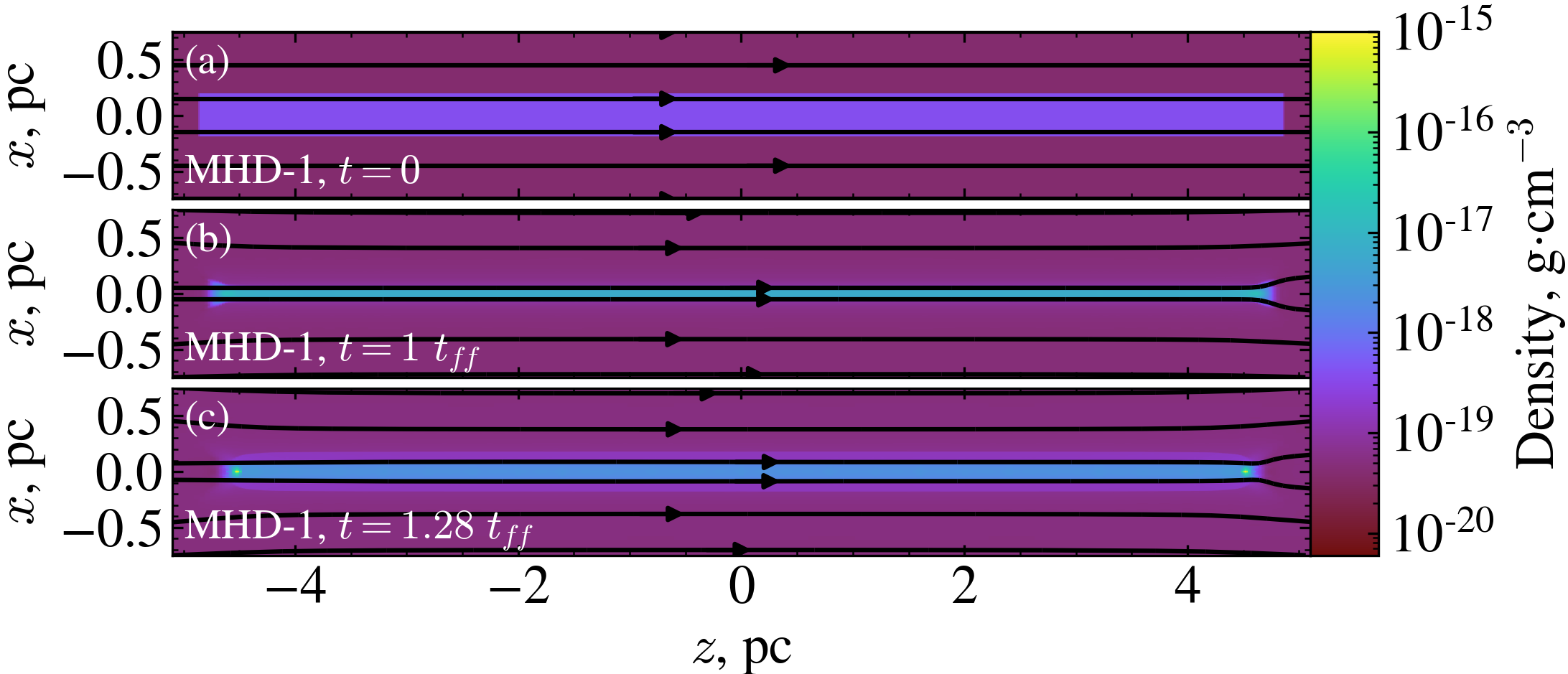}
\caption{The distribution of gas density (color map) and magnetic field lines (black lines with arrows) of the filament in the $x-z$-plane for run MHD-1 at $t = 0$~(a), $t = 1 \, t_{\rm ff}$~(b), $t = 1.28 \, t_{\rm ff}$~(c).}
\label{Fig. 2:}
\end{figure}

In Figure \ref{Fig. 2:}, we plot the gas density distribution in the $x-z$~plane for run 'MHD-1' at $t = 0,\, 1 \, t_{\rm ff},\, 1.28 \, t_{\rm ff}$. Figure 2 shows that the magnetic filament collapses to radius~$r \approx 0.1$~pc and density $n \approx 1.77 \cdot 10^{8}$~cm$^{-3}$ by the time of $t = 1 \, t_{\rm ff}$. After that, the effective adiabatic index increases from $1$~to $2$~due to the action of the electromagnetic force, the collapse along $r$~is stopped by the magnetic pressure gradient, and the cloud starts to oscillate along the radius. Two cores with density $n = 1.7 \cdot 10^8$~cm$^{-3}$ form at the edges of the cloud by the time $t = 1.28 \, t_{\rm ff}$. The properties of the cores are discussed further in the section "Characteristics of forming cores".

\subsection{Influence of the Magnetic Field on the Evolution}
\label{sect:mag_results}

\begin{figure}[h!]
\includegraphics[width=0.9\textwidth]{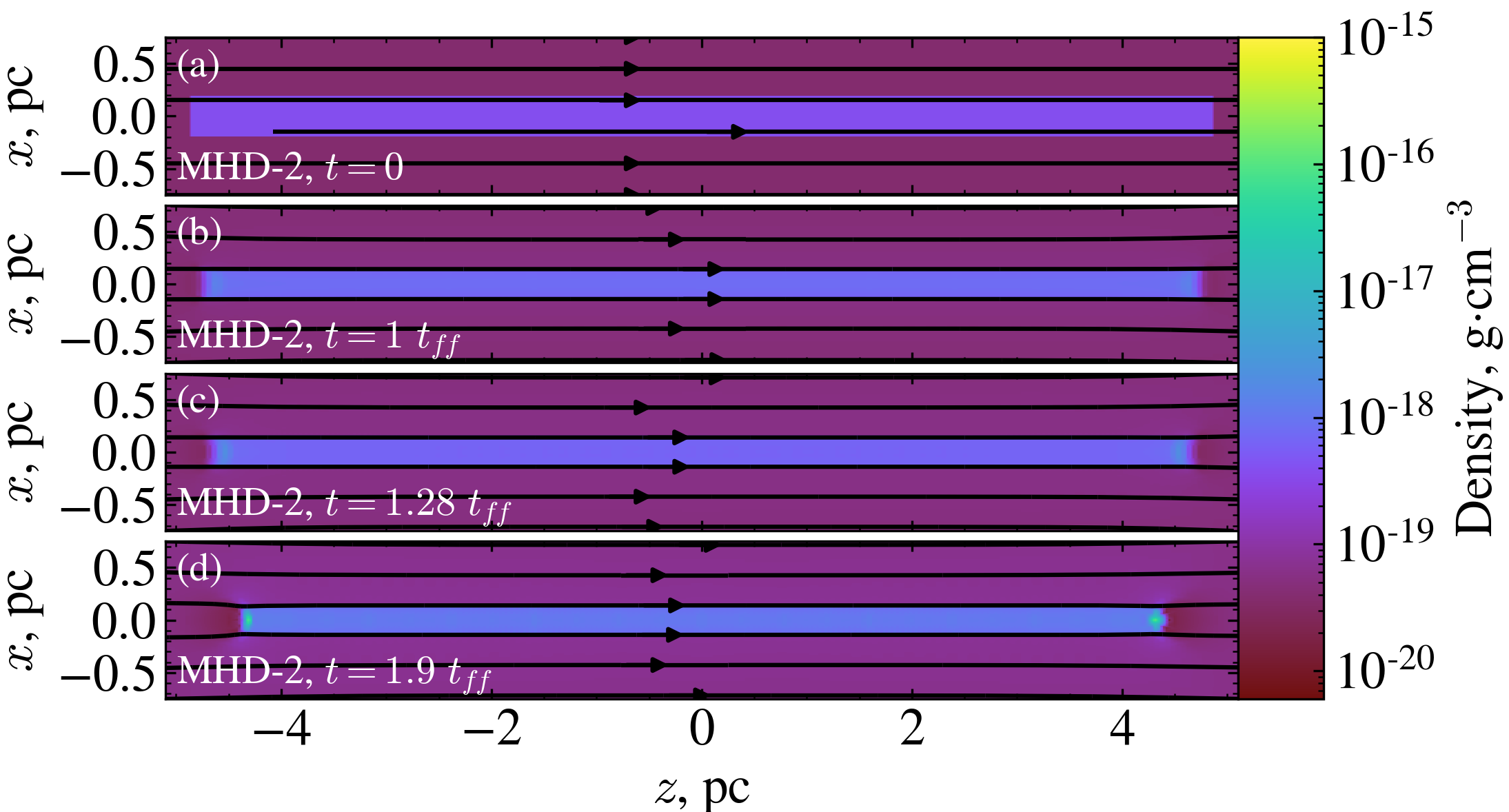}
\caption{The distribution of gas density (color map) and magnetic field lines (black lines with arrows) of the filament in the $x-z$-plane for run MHD-2 at $t = 0$~(a), $t = 1 \, t_{\rm ff}$~(b), $t = 1.28 \, t_{\rm ff}$~(c), and $t = 1.9 \, t_{\rm ff}$~(d).}
\label{Fig. 3:}
\end{figure}

In Figure \ref{Fig. 3:}, we plot the gas density distribution in the $x-z$-plane for run 'MHD-2' at $t = 0, \, 1 \, t_{\rm ff}, \, 1.28 \ t_{\rm ff}, \, 1.9 \, t_{\rm ff}$. In this run, the picture of the collapse is qualitatively similar to the results for run 'MHD-1'. The filament collapses to radius of $r \approx 0.3$~pc and concentration of $n \approx 2 \cdot 10^{5}$~cm$^{-3}$ at time $1 \, t_{\rm ff}$, after which the collapse stops and the filament oscillates along the radius. The density of the cores formed at the ends of the filament is $n = 2 \cdot 10^7$~cm$^{-3}$ at the time $1.9 \, t_{\rm ff}$.
 
\begin{figure}[t]
\includegraphics[width=0.9\textwidth]{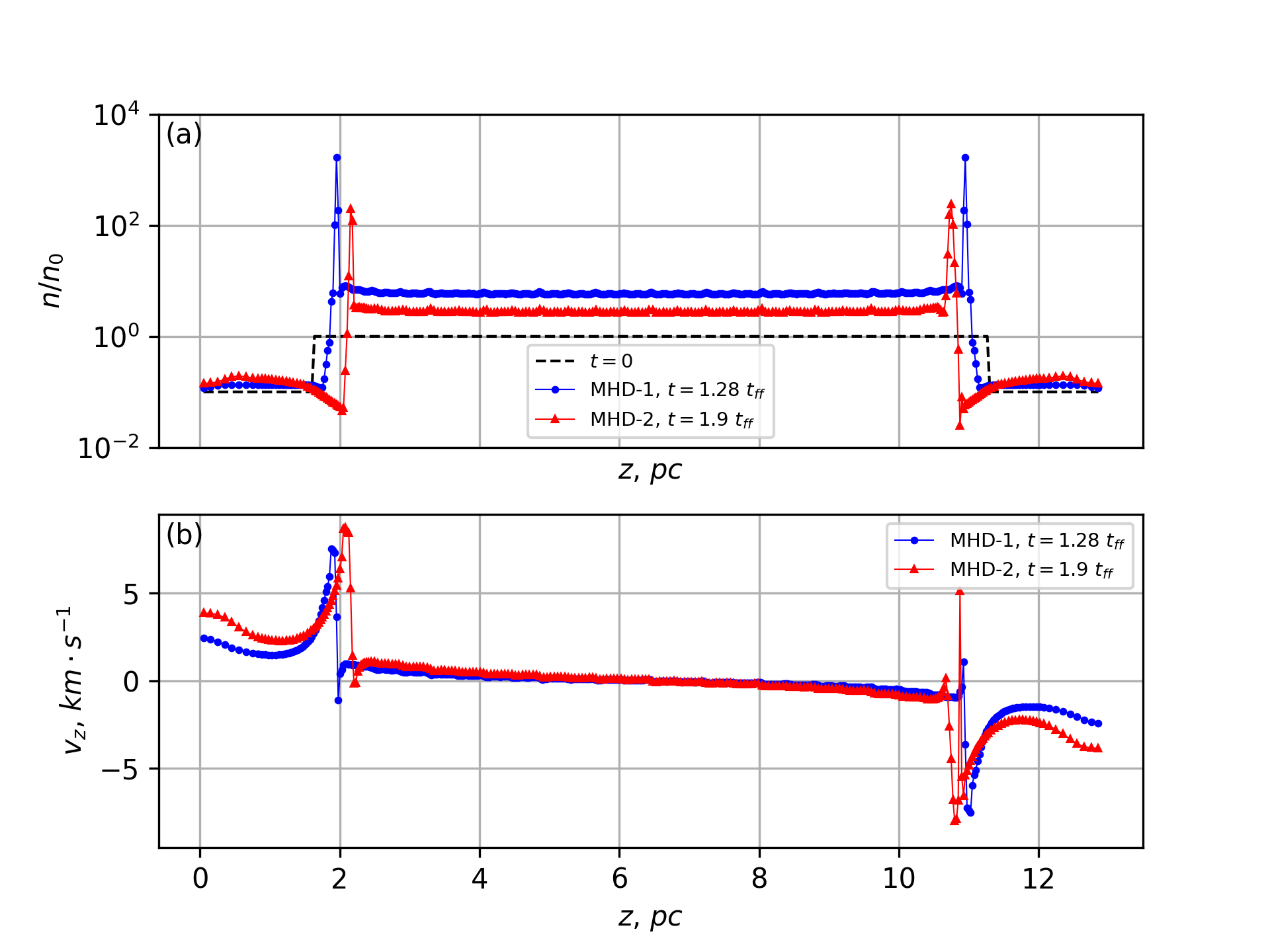}
\caption{(a) The profiles of density along the $z$-axis for MHD runs at , $t = 0, \, 1.28 \, t_{\rm ff}$~and $t = 1.9 \, t_{\rm ff}$ . (b) corresponding velocity profiles.}
\label{Fig. 4:}
\end{figure}

In Figure \ref{Fig. 4:}, we show the evolution of the density and velocity profiles along the filament's axis in MHD runs. The filament edges are located at $z_{\rm L} = 1.6$~pc and $z_{\rm R} = 11.3$~pc initially. Figure 4 shows that, in run MHD-1, peaks densities $n = 1.7 \cdot 10^{8}$~cm$^{-3}$ and velocities $v_{z} = 3.6$~km$\cdot$s$^{-1}$ are observed at the ends of the filament ($z_{\rm L} = 1.9$~pc and $z_{\rm R} = 11$~pc) at the moment of time $t = 1.28 \, t_{\rm ff}$. These peaks correspond to the cores formed at the ends of the filament. The speed of the left core is positive, the speed of the right one is negative. Corresponding Mach number is $M = 6$, that is, the cores move at supersonic speeds towards each other.

In run MHD-2, the densities and velocities of the cores located at $z_{\rm L} = 2.15$~pc and $z_{\rm R} = 10.75$~pc at the time moment are equal to $n = 2 \cdot 10^{8}$~cm$^{-3}$~and $|v_{z}|=5$~km$\cdot$s$^{-1}$, respectively, and the corresponding Mach number is equal to $M = 8.3$.

\subsection{Characteristics of Forming Cores}
\label{sect:cores}

\begin{figure}[t]
\includegraphics[width=0.9\textwidth]{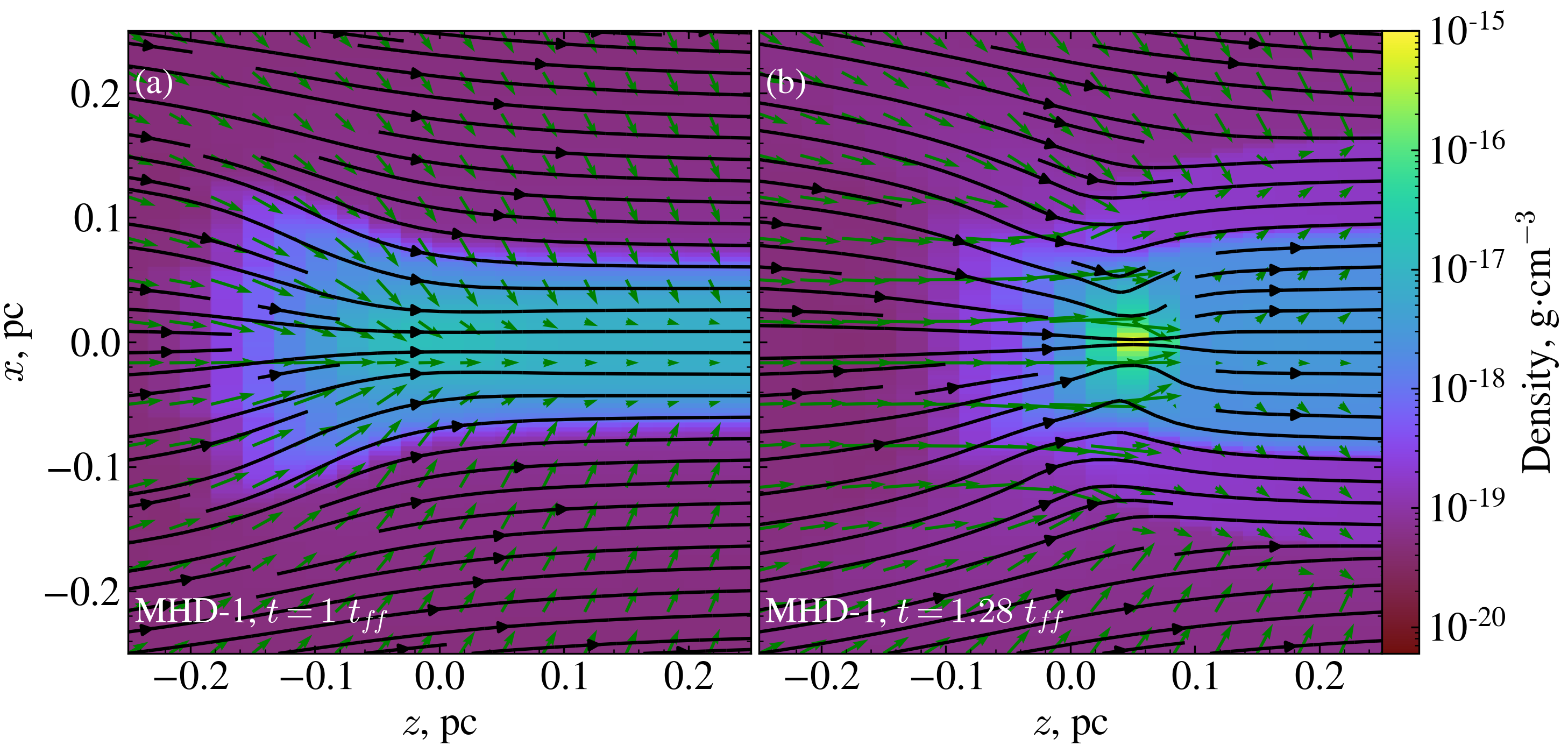}
\caption{The distribution of gas density (color map), velocity field (green arrows) and magnetic field lines (black lines with arrows) in core formation region for run MHD-1 at $t = 1 \, t_{\rm ff}$~(a) and $t = 1.28 \, t_{\rm ff}$~(b).}
\label{Fig. 5:}
\end{figure}

\begin{figure}[t]
\includegraphics[width=0.9\textwidth]{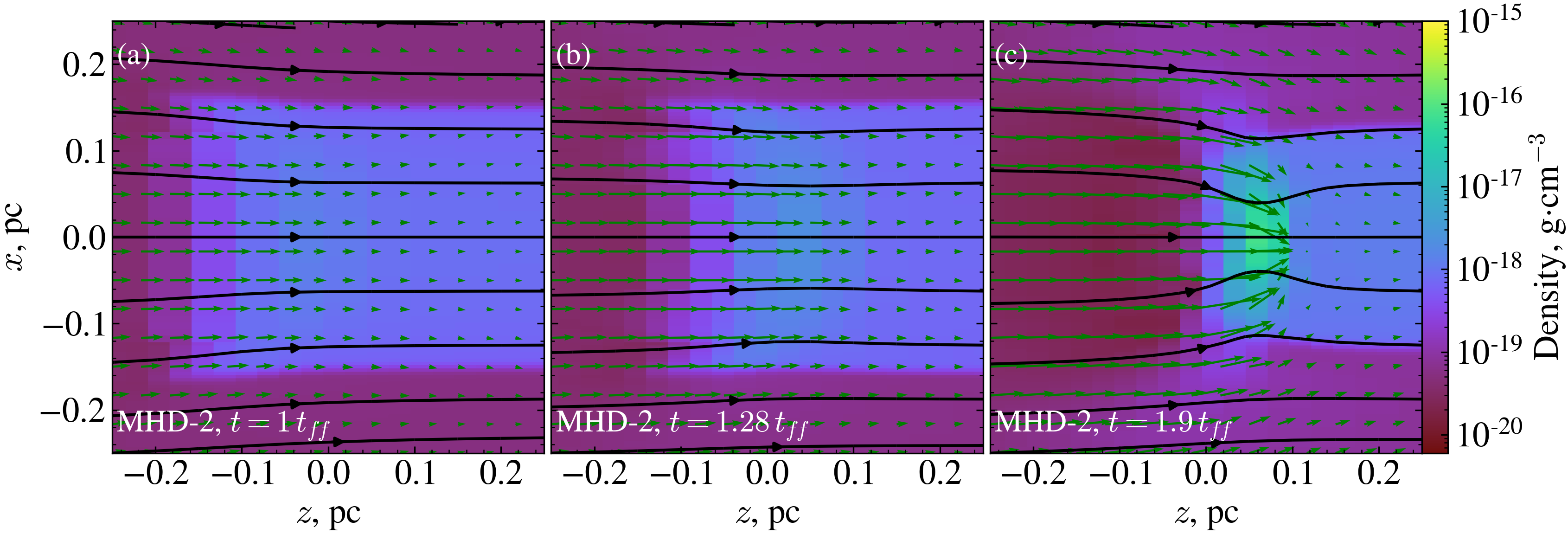}
\caption{The distribution of gas density (color map), velocity field (green arrows) and magnetic field lines (black lines with arrows) in core formation region for run MHD-2 at $t = 1 \, t_{\rm ff}$~(a), $t = 1.28 \, t_{\rm ff}$~(b) and $t = 1.9 \, t_{\rm ff}$~(c).}
\label{Fig. 6:}
\end{figure}

In Figure \ref{Fig. 5:}, we plot the distribution of gas density, magnetic field lines and velocity field for run MHD-1 in the region of core formation in the plane. Two moments of time are presented: $t = 1$~and $1.28 \, t_{\rm ff}$ . Due to symmetry, only the core located at the "left" end of the filament is shown. In \ref{Fig. 6:}, we plot similar distribution for run MHD-2 at moments of time $t = 1, \, 1.28 \, t_{\rm ff}$~and $1.9 \, t_{\rm ff}$~in the $x-z$-plane
 
Table~\ref{Table I:} gives the following characteristics of the cores formed in MHD runs: sizes along the filament radius $r$~and the $z$-axis (columns 3 and 4), concentration $n$~(column 5), mass $M$~(column 6) and velocity $v_{z}$~(column 7). The table shows that the cores of larger radius $d_{r}$ and lower density are formed in the simulation with larger magnetic field strength. This is due to the fact that the influence of the magnetic pressure gradient on the dynamics of the filament is increased in the case of stronger magnetic field. The sizes of the cores along the $z$-axis do not depend on the strength of the initial magnetic field, since the parallel magnetic field does not prevent collapse of the filament along its axis. The masses of the cores grows as they move and reach $12.5\,M_\odot$ in run MHD-1 and $23\,M_\odot$ in run MHD-2.

\begin{table}[t]
\caption{Characteristics of the cores in MHD runs}
\begin{tabular}{ccccccc}
\hline
Run & Time, $t_{\rm ff}$ & $d_{r}$, pc & $d_{z}$, pc & $n$, cm$^{\rm -3}$ & $M$, $M_{\odot}$ &$v_{z}$, km$\cdot$s$^{-1}$\\
  (1) & (2) & (3) & (4) & (5) & (6) & (7) \\
\hline
 MHD-1& 1.28 & 0.0075 & 0.025 & $1.7\cdot10^{8}$ & $12.5$ &$3.6$\\
\multirow{2}{*}{MHD-2} & 1.28 &  $0.06$ & $0.025$ & $5.3\cdot10^{5}$ & $2.5$ & $2.7$\\
 & 1.9 & $0.03$ & $0.025$ & $2\cdot10^{7}$ & 23.6& $5.3$\\
\hline
\label{Table I:}
\end{tabular}

\end{table}

\section*{Summary}
\label{sect:outro}

The influence of the magnetic field on the evolution of molecular filament and the characteristics of
the cores formed in the filament as a result of end dominated collapse is investigated in this work. For
this purpose, we performed a set of numerical MHD simulations of the gravitational collapse of a cylindrical molecular filament with different values of the parallel magnetic field.

Our simulations confirm the conclusions of earlier studies that the filament without magnetic field collapses freely along its radius. Fragmentation of the filament and the formation of the cores at its ends do not occur during the collapse, since the collapse along the radius occurs on a shorter time scale.

The magnetic pressure gradient prevents collapse and leads to decaying oscillations of the filament along
its radius. During the evolution of magnetic filaments, dense clumps (cores) form at the ends of the filament due to the gravitational focusing. The cores move towards the center of the cloud at supersonic speeds of $3$~to $5$~km$\cdot$s$^{-1}$. The clouds with stronger magnetic field are characterized by the formation of the cores of larger sizes and lower density, since the influence of the magnetic field pressure gradient along the radius increases with increasing magnetic field strength. The masses of the cores increase during the evolution of the filament and lie in the range $\approx 10-20\,M_\odot$.

Our simulations indicate that the end-dominated collapse is a natural result of the evolution of the filaments with parallel magnetic field. It can be assumed that the observed filaments with clumps located at ends (e.g.,~\cite{dewanghan}) are supported from gravitational fragmentation by parallel magnetic field. Additional support against gravity can be provided by turbulence within the filament~\cite{sw2015, federrath2021}.

We plan to further develop the presented model and model the evolution of initially non-uniform filaments taking into account their rotation and/or internal turbulence. It is of particular interest to study the
fragmentation of the filament with parallel magnetic field due to the gravitational instability as described by the theory of Chandrasekhar and Fermi~\cite{ChFe}.

\begin{acknowledgments}
The work is financially supported by the Foundation for Perspective Research of the Chelyabinsk State University (project 2023/7). The work by S.A. Khaibrakhmanov is supported by the Russian Ministry of Science and Higher Education via the Project FEUZ-2020-0038. The simulations were carried out using the computational cluster of the Chelyabinsk State University.
\end{acknowledgments}


\begin{thebibliography}{99}
\bibitem[{Andre} {et.~al.}(2014)]{andre2014}
{Andre} P., {Di Francesco} J., {Ward-Thompson} D., {Inutsuka} S.~I., {Pudritz} R.~E., {Pineda} J.~E., 2014, Protostars and Planets VI, 27

\bibitem[{Dudorov} \& {Khaibrakhmanov}(2017)]{DK2017}
{Dudorov} A.~E., {Khaibrakhmanov} S.~A. 2017, Open Astronomy, 26, 285

\bibitem[{Konyves} {et.~al.}(2015)]{konyves2015}
{Konyves} V.,  {Andre} Ph.,  {Men'shchikov} A.,  {Palmeirim} P., {Arzoumanian} D., {Schneider} N., {Roy} A., {Didelon} P., {Maury} A., {Shimajiri} Y., {Di Francesco} J., {Bontemps} S., {Peretto} N., {Benedettini} M., {Bernard} J.~Ph., {Elia} D., {Griffin} M.~J., {Hill} T., {Kirk} J., {Ladjelate} B., 2015, Astronomy and Astrophysics, 584, 33

\bibitem[{Ward-Thompson} {et.~al}(2017)]{ward_thompson2017}
{Ward-Thompson} D.,  {Pattle} K., {Bastien} P., {Furuya} R.~S., {Kwon} W., {Lai} S.~P., {Qiu} K., {Berry} D., {Choi} M., {Coude} S., {Di Francesco} J., {Hoang} T., {Franzmann} E., {Friberg} P., {Graves} S.~F., {Greaves} J.~S., {Houde} M., {Johnstone} D., {Kirk} J.~M., {Koch} P.~M. 2017, The Astrophysical Journal, 842, 10

\bibitem[{Hacar} {et.~al.}(2023)]{hacar2023}
{Hacar} A., {Clark} S.~E., {Heitsch} F., {Kainulainen} J., {Panopoulou} G.~V, {Seirfried} D., {Smith} R., 2023, Protostars and Planets VII, ASP Conference Series, Proceedings of a conference, 534, 153

\bibitem[{Bastien}(1983)]{Bastien}
{Bastien} P., 1983, Astronomy and Astrophysics, 119, 109 

\bibitem[{Dewanghan} {et.~al}(2019)]{dewanghan}
{Dewangan} L.~K, {Pirogov} L.~E., {Ryabukhina} O.~L., {Ojha} D.~K., {Zinchenko} I., 2019, The Astrophysical Journal, 877, 1

\bibitem[{Chandrasekhar} \& {Fermi}(1953)]{ChFe}
{Chandrasekhar} S., {Fermi} E., 1953, Astrophysical journal, 118, 116

\bibitem[{Stodolkiewicz} (1963)]{stod}
{Stodolkiewicz} J.~S, 1963, Acta Astronomica, 13, 30

\bibitem[{Ostriker} (1964)]{ostriker}
{Ostriker} J., 1964, Astrophysical Journal, 140, 1056

\bibitem[{Inutsuka} \& {Miyama}(1997)]{IM}
{Inutsuka} S., {Miyama} S.~M., 1997, 480, 681

\bibitem[{Shimajiri} {et.~al}(2023)]{shimajiri}
{Shimajiri} Y., {Andre} Ph., {Peretto}, {Arzoumanian} D., {Ntormousi} E., {Konyves} V., 2023, Astronomy and Astrophysics, 627, 1

\bibitem[{Ryabukhina} {et.~al.}(2022)]{ryabukhina2022}
{Ryabukhina} O.~L., {Kirsanova} M.~S., {Henkel} C., {Wiebe} D.~C, 2022, Mon. Not. R. Astr. Soc., 517, 4669

\bibitem[{Seifried} \& {Walch}(2015)]{sw2015}
{Seifried} D., {Walch} S., 2015, Mon. Not. R. Astr. Soc., 452, 2410

\bibitem[{Dudorov} \& {Khaibrakhmanov}(2014)]{DK2014}
{Dudorov} A.~E., {Khaibrakhmanov} S.~A. 2014, Astrophysics and Space Science, 352, 103

\bibitem[{Dudorov} \& {Sazonov}(1987)]{DS1987}
{Dudorov} A.~E., {Sazonov} Yu.~V. 1987, Nauchnye Informatsii, 63, 68

\bibitem[{Fryxell} {et.~al.}(2018)]{Fryxell}
{Fryxell} B., {Olson} K., {Ricker} P. et.~al. 2000, Astrophysical journal Supplement Series, 131, 273

\bibitem[{van Leer} (1979)]{muscl}
{van Leer} B., 1979, JCP, 32, 101

\bibitem[{Barnes} \& {Hut}(1986)]{BHtree}
{Barnes} J., {Hut} P., 1986, Nature, 326, 446

\bibitem[{Federrath} {et.~al.}(2021)]{federrath2021}
{Federrath} C., {Klessen} R.~S., {Iapichino} L., {Beattie} J.~R., 2021, Nature Astronomy, 5, 365


\end{thebibliography}
\end{document}